\documentstyle[preprint,aps]{revtex}

\begin{document}

\draft

\title{New Limits on the Couplings of Light Pseudoscalars from Equivalence
Principle Experiments}

\author{Ephraim Fischbach$^{(1)}$ and Dennis E. Krause$^{(2)}$}

\address{$^{(1)}$Physics Department, Purdue University, West Lafayette, IN 
47907-1396}

\address{$^{(2)}$Physics Department, Wabash College, Crawfordsville, IN 
47933-0352}

\date{\today}

\maketitle

\begin{abstract}

The exchange of light pseudoscalar quanta between fermions leads to
long-range spin-dependent forces in order $g^2$, where $g$ is the 
pseudoscalar-fermion coupling constant.  We demonstrate that
laboratory bounds on the Yukawa couplings
of pseudoscalars to nucleons can be significantly improved
 using results
from recent equivalence principle experiments, which are
sensitive to the spin-independent long-range forces that arise in order
$g^4$ from two-pseudoscalar exchange.

\end{abstract}


%\begin{multicols}{2}
\pagebreak

	  It is well known that the exchange of a light pseudoscalar
quantum $(\phi)$ with mass $m$ between two fermions
$(\psi)$ of mass $M$ gives rise to a long-range spin-dependent
fermion-fermion interaction.  If we describe the
fundamental coupling via the usual Lagrangian density
\begin{equation}
 {\cal L} (x) = ig \bar{\psi}(x)\gamma_5 \psi(x) \phi(x),
\label{L}
\end{equation}
where $g$ is the pseudoscalar coupling constant, then the
spin-dependent  potential between two identical spin-1/2
fermions is given by \cite{Bohr and Mottelson}
\begin{eqnarray}   
   V^{(2)} \left(\vec{r};\vec{\sigma}_1,\vec{\sigma}_2\right) & = &
      \frac{g^{2}}{16\pi M^{2}}
          \left\{(\vec{\sigma}_{1}\cdot\hat{r})(\vec{\sigma}_{2}\cdot\hat{r})
      \left[\frac{m^{2}}{r} + \frac{3m}{r^{2}} +
\frac{3}{r^{3}}
    \right]   \right.
   \nonumber  \\
   &    & \left. {} - (\vec{\sigma}_{1}\cdot\vec{\sigma}_{2})
        \left[\frac{m}{r^{2}} + \frac{1}{r^{3}}\right]\right\}e^{-mr}.
\label{massive V_ps}
\end{eqnarray}
Here $r = |\vec{r}| = |\vec{r}_1 - \vec{r}_2|$ is the distance
between fermions 1 and 2, 
$(1/2)\vec{\sigma}_{1,2}$ are the fermion spins ($\hbar = c = 1$),
and we have dropped a term proportional to $\delta^{3}(r)$.   Our focus
in this paper will be on the
$m = 0$ limit \cite{Anselm} of Eq.~(\ref{massive V_ps}), which
characterizes the long-range interaction between fermions when
$1/m$ is large compared to the size of the apparatus, 
\begin{mathletters}
\begin{equation}
V^{(2)} (\vec{r};\vec{\sigma}_1, \vec{\sigma}_2)
     \stackrel{m = 0}{\longrightarrow}
\frac{g^{2}}{16\pi M^{2}}
\frac{S_{12}}{r^{3}},
\label{massless V_ps}
\end{equation}
\begin{equation}
S_{12} \equiv 3(\vec{\sigma}_{1}\cdot\hat{r})(\vec{\sigma}_{2}\cdot\hat{r}) -
        (\vec{\sigma}_{1}\cdot\vec{\sigma}_{2}).
\end{equation}
\end{mathletters}
Limits on 
$g^2/4\pi$ derived from recent spin-dependent experiments are
summarized by Ritter, {\em et al.} \cite{Gillies}.  Although these
limits appear at first to be quite restrictive, they are not nearly as
stringent as the limits implied by recent spin-independent tests of the
equivalence principle, which also probe for the presence of new
long-range forces. For example, if the coupling of a new long-range
vector field $A_\mu$ to fermions is described by the Lagrangian
\begin{equation}
  {\cal L} = if~\bar{\psi}(x) \gamma_\mu \psi(x) A_\mu(x),
\end{equation}
then typical limits on $f^2/4\pi$ over laboratory distance
scales are $f^2/4\pi \lesssim 10^{-46}$
\cite{Gundlach,Fischbach book} compared to $g_{e}^{2}/4\pi 
\lesssim 10^{-16}$, where $g_{e}$ is the pseudoscalar coupling
to electrons \cite{Gillies}.  Among the reasons for the differing
sensitivities of spin-dependent and spin-independent experiments are
\cite{Fischbach book}:  1) The strength of the spin-dependent coupling
in Eq.~(\ref{massless V_ps}) is suppressed relative to that for the
spin-independent coupling by a factor of order $1/(MR)^2$,
where $R$ is the characteristic size of the experimental
apparatus.  If $M$ denotes the electron mass and
$R = 1$ m, then $1/(MR)^2 \simeq 1.5 \times 10^{-25}$.
2)  Test masses which have a net electron-spin polarization
can also interact electromagnetically. 
Since the electromagnetic background is many orders of magnitude
larger than the effects expected from a putative new force,
special materials (such as Dy$_{6}$Fe$_{23}$)
 and methods must be used which limit the
sizes of the samples that can be studied.  
3) Furthermore, even in these special materials, only a small
fraction of the test masses actually contributes, since the
net polarization is only 0.4 electrons per Dy$_{5}$Fe$_{23}$
molecule \cite{Ritter}.
4) The spin-dependent
couplings of light pseudoscalars to 
nucleons  are further suppressed by the dilution
of the electron polarization as it is transferred to the
nucleons.

The disparity in the limits set on $g^2$ and $f^2$, by
spin-dependent and spin-independent experiments respectively,
raises the question of whether interesting limits on
$g^2$ can also be inferred from spin-independent searches
for macroscopic forces.  The exchange of two pseudoscalars,
as shown in Fig. 1, gives rise to a spin-independent
potential $V^{(4)}(r)$ in order $g^4$ which has been calculated
by a number of authors \cite{Drell,Ferrer}.  In the limit 
$m \rightarrow 0,  ~V^{(4)}(r)$ is given by
\begin{equation}
 V^{(4)} (r) = -\frac{g^4}{64\pi^3M^2} \frac{1}{r^3}
\equiv  g^{4}f(r).
\label{V_ps 4}
\end{equation}
Interestingly, $V^{(4)}$ and $V^{(2)}$ have the same
functional dependence on $M$ and $r$ in the $m = 0$ limit,
and the ratio of their strengths (per pair of interacting
particles) is
\begin{equation}
 \frac{|V^{(2)}(r;\vec{\sigma}_1,\vec{\sigma}_2)|}
         {|V^{(4)} (r)|}
     = \frac{4\pi^2 |\langle S_{12}\rangle|}
            {g^2}~~,
\label{V ratio}
\end{equation}
where $\langle S_{12}\rangle$ is determined by averaging over the
polarizations of samples 1 and 2.  We see from Eq.~(\ref{V ratio}) that
although
$V^{(4)}$ is suppressed relative to $V^{(2)}$ by the factor
$g^2/4\pi^{2}$, $V^{(2)}$ is suppressed relative to $V^{(4)}$ by the
factor $\langle S_{12}\rangle$.  Moreover, $V^{(2)}$ is further suppressed
relative to $V^{(4)}$ by virtue of the fact that there are fewer
contributions to $\sum V_{ij}^{(2)}$ than to
$\sum V_{ij}^{(4)}$, since the source masses are necessarily
smaller in the spin-dependent experiments.  

As we show in  the ensuing discussion, the net effect of the 
various suppression factors in
Eq.~(\ref{V ratio}) is that the most stringent laboratory limits on
Yukawa couplings of pseudoscalars to protons,
neutrons, (and ultimately quarks) arise from spin-independent
equivalence  principle experiments which
constrain $V^{(4)}$, rather than from spin-dependent 
experiments which are sensitive to
$V^{(2)}$.  Since the couplings of axions to
fermions involve derivatives, the resulting 2-axion potential
varies as $1/r^{5}$ rather than as $1/r^{3}$, as has been noted
by Ferrer and Grifols \cite{Ferrer}.  Hence, the numerical
results of the present paper do not apply to axions directly,
although the present formalism can be taken over for axions with
appropriate modifications.

Consider the interaction between two objects 1 and 2 containing
$N_{1}$ ($Z_{1}$) neutrons (protons), and $N_{2}$ ($Z_{2}$) neutrons
(protons), respectively.  The total energy $W$ is obtained by summing
the pairwise interactions arising from Eq.~(\ref{V_ps 4}) after
replacing the generic coupling constant $g^{4}$ by $g_{n}^{4}$,
$g_{p}^{4}$, or $g_{p}^{2}g_{n}^{2}$ for n-n, p-p, and n-p
interactions, respectively.  Here $g_{n}$ ($g_{p}$) denotes the
pseudoscalar coupling constant appearing in Eq.~(\ref{L}) when
$\psi$ is a neutron (proton).  From Eq.~(\ref{V_ps 4}) $W$ can be
expressed in the form
\begin{equation}
W = [g_{p}^{4}Z_{1}Z_{2} + g_{n}^{4}N_{1}N_{2} +
      g_{n}^{2}g_{p}^{2}(Z_{1}N_{2} + Z_{2}N_{1})]
       \langle f(r)\rangle,
\label{W}
\end{equation}
where $\langle f(r)\rangle$ is obtained from Eq.~(\ref{V_ps 4}) by
integrating over the mass distributions of the two objects.

In a typical equivalence principle experiment object 1 is an extended
source toward which the relative accelerations of samples 2 and $2'$
(with masses $M_{2}$ and $M_{2'}$)
are being measured.  If the dimensions of the test masses are small
compared to the size of the source, the force $\vec{F}(\vec{r}\,)$
exerted by the source on test mass 2 (located at $\vec{r}\,$) can be
written in the form
\begin{mathletters}
\begin{equation}
\vec{F}(\vec{r}\,) = [g_{p}^{4}Z_{1}Z_{2} + g_{n}^{4}N_{1}N_{2}
           + g_{n}^{2}g_{p}^{2}(Z_{1}N_{2} + Z_{2}N_{1})]
 \vec{\cal F}(\vec{r}\,),
\label{F}
\end{equation}
\begin{equation}
\vec{\cal F}(\vec{r}\,) = \left(\frac{-3}{64\pi^{3}M^{2}{\cal
V}_{1}}\right)
\int d^{3}r_{1}'\, \frac{(\vec{r} - \vec{r}_{1}{}'\,)}
   {|\vec{r} - \vec{r}_{1}{}'\,|^{5}},
\end{equation}  
\end{mathletters}
where ${\cal V}_{1}$ is the volume of the source.  It
follows from Eq.~(\ref{F}) that the experimentally measured
acceleration difference $\Delta\vec{a}_{2-2'} \equiv \vec{a}_{2} -
\vec{a}_{2'}$ is given by
\begin{equation}
\Delta\vec{a}_{2-2'} = 
   \vec{\cal F}(\vec{r}\,)
    \left(\frac{M_{1}}{m_{H}^{2}}\right)
    \left[g_{p}^{2}\left(\frac{Z_{1}}{\mu_{1}}\right)
         + g_{n}^{2}\left(\frac{N_{1}}{\mu_{1}}\right)
     \right]
   \left[g_{p}^{2}\Delta\left(\frac{Z}{\mu}\right)_{2-2'}
  + g_{n}^{2}\Delta\left(\frac{N}{\mu}\right)_{2-2'}\right],
\label{Delta a}
\end{equation}
where $M_{1}$ is the source mass, $\Delta(Z/\mu)_{2-2'} = Z_{2}/\mu_{2}
- Z_{2'}/\mu_{2'}$, etc.,
$\mu_{i} = M_{i}/m_{\rm H}$, and $m_{\rm H}~=~m(_{1}{\rm
H}^{1})$  
\cite{Fischbach book}.
Except for $g_{p}^{2}$ and $g_{n}^{2}$, the right-hand side of
Eq.~(\ref{Delta a}) is known, and hence an experimental determination of
$\Delta\vec{a}_{2-2'}$ leads to a constraint on $g_{p}^{2}$ and
$g_{n}^{2}$.

Examination of Eq.~(\ref{Delta a}) leads to the observation that there
are two classes of constraints on $g_{p}^{2}$ and $g_{n}^{2}$,
depending on the relative signs of $\Delta(Z/\mu)_{2-2'}$ and
$\Delta(N/\mu)_{2-2'}$.  Since $g_{p}^{2}$, $g_{n}^{2}$, $N_{1}$, and
$Z_{1}$ are all inherently positive, the right-hand side of
Eq.~(\ref{Delta a}) cannot vanish if $\Delta(Z/\mu)_{2-2'}$ and 
$\Delta(N/\mu)_{2-2'}$ have the same sign, unless $g_{p}^{2}$ and
$g_{n}^{2}$ themselves do.  It follows that in this circumstance an
experimental bound on $\Delta\vec{a}_{2-2'}$ leads to an absolute
upper bound on either $g_{p}^{2}$ or $g_{n}^{2}$.  We refer to such
constraints as ``elliptical'', since Eq.~(\ref{Delta a}) produces
ellipses in the $x$-$y$ plane defined by $x = g_{p}^{2}$ and $y =
g_{n}^{2}$.  By contrast, if $\Delta(Z/\mu)_{2-2'}$ and
$\Delta(N/\mu)_{2-2'}$ have opposite signs, the right-hand side of
Eq.~(\ref{Delta a}) can vanish whenever $g_{p}^{2}$ and $g_{n}^{2}$
satisfy
\begin{equation}
\frac{g_{p}^{2}}{g_{n}^{2}} =
      - \frac{\Delta(N/\mu)_{2-2'}}{\Delta(Z/\mu)_{2-2'}},
\label{no limit condition}
\end{equation}
and hence $g_{p}^{2}$ and $g_{n}^{2}$ can be arbitrarily large and
still be compatible with any experimental bound on
$\Delta\vec{a}_{2-2'}$.  We term such constraints ``hyperbolic'',
since in this case Eq.~(\ref{Delta a}) leads to hyperbolas in the
$x$-$y$ plane.  The asymptotes of these hyperbolas in the
(physical) first quadrant lie near the line $y = x$, which represents
the locus of points satisfying Eq.~(\ref{no limit condition})
\cite{Long paper}.

It is instructive to contrast the constraints arising from
$V^{(4)}$ in Eq.~(\ref{V_ps 4}) with those arising in second
order from the exchange of a scalar or vector field, as in the usual
``fifth force'' scenario \cite{Fischbach book}.  The expression for
$\Delta\vec{a}_{2-2'}$ in this case has the same general form as in
Eq.~(\ref{Delta a}) except that $g_{p,n}^{2} \rightarrow g_{p,n}$.
Since $g_{p}$ and $g_{n}$ can each be positive or negative, no choice
of samples 2 and $2'$ can ensure that the coefficient of $\vec{\cal
F}(\vec{r})$ will have a unique sign, and hence there are no
elliptical constraints in the conventional ``fifth force'' case.  Note
that $\Delta\vec{a}_{2-2'}$ can vanish not only when the analog of
Eq.~(\ref{no limit condition}) holds for the test masses, but also
when the source strength vanishes as happens when $(g_{p}Z_{1} +
g_{n}N_{1}) = 0$ \cite{Sensitivity}.  It follows from this discussion
that the novel feature of $V^{(4)}$ is that it gives rise to
elliptical constraints, and hence to absolute bounds on $g_{p}^{2}$
and $g_{n}^{2}$, for appropriate choices of 2 and $2'$.  To determine
which pairs of elements would produce elliptical constraints, we have
evaluated $\Delta(Z/\mu)$ and $\Delta(N/\mu)$ for the 4,186 pairs that
can be formed from the first 92 elements, and found 7 possible pairs:
He-N, He-O, N-O, S-Ca, Br-Mo, Li-Ru, and Pt-Rn.  Among these, Li-Ru is
the most obvious choice, where the Li sample could be gold-plated to
prevent oxidation.  Other choices involving compounds are also
possible, as we discuss in greater detail elsewhere \cite{Long
paper}.

As we now demonstrate, if the preceding formalism is combined with the
recent results of Gundlach, {\em et al.} \cite{Gundlach}, the laboratory
limits on $g_{p}^{2}$ and $g_{n}^{2}$ can be significantly improved.
This experiment compared the accelerations of  test bodies composed of
Cu and a Pb alloy toward a 2620 kg depleted U source, and they found for
the acceleration difference
\begin{equation}
\Delta\vec{a}_{2-2'} = \vec{a}_{\rm Cu} - \vec{a}_{\rm Pb}
   = \hat{r}(-0.7 \pm 5.7)\times 10^{-13}\,\,\,\mbox{cm/s$^{2}$},
\label{Gundlach's Delta a}
\end{equation}
where $\hat{r}$ is a unit vector in the direction of the field
$\vec{\cal F}$ produced by the source.  Since the U source was
positioned close to the test masses, this experiment can be used to
set limits on short-range  interactions of the form $V(r) =
\Lambda_{N}(r_{0}/r)^{N-1}(\hbar c/r)$, with $N = 3$ corresponding to
$V^{(4)}$ in Eq.~(\ref{V_ps 4}).  Combining Eq.~(\ref{Delta
a}) with the bound from Ref.~\cite{Gundlach}, $\Lambda_{3} < 6 \times
10^{-16}$, leads to the constraint
\begin{equation}
\frac{|\Delta\vec{a}_{2-2'}|}{(\mbox{1 cm/s$^{2}$})}
 = (9.6 g_{p}^{2} + 15.3 g_{n}^{2})
    |g_{p}^{2}\Delta(Z/\mu)_{2-2'} + g_{n}^{2}\Delta(N/\mu)_{2-2'}|,
\label{Gundlach's limit}
\end{equation}
which applies to any test masses 2 and $2'$ in Ref.~\cite{Gundlach}.
For the actual samples used, 2~=~Cu and $2'$~=~Pb alloy,
$\Delta(Z/\mu)_{2-2'} = 0.05925$, $\Delta(N/\mu)_{2-2'} = -0.05830$,
and the slope of the asymptote for the hyperbolic constraint implied by
Eq.~(\ref{Gundlach's limit}) is 0.05925/0.05830 = 1.016.  Inserting
these results for 2 and $2'$ into Eq.~(\ref{Gundlach's limit}) along
with the $1\sigma$ bound in Eq.~(\ref{Gundlach's Delta a}),
$|\Delta\vec{a}_{2-2'}|<~6.4\times~10^{-13}\,\,\mbox{cm/s$^{2}$}$,
leads to the final result,
\begin{equation}
(9.6 g_{p}^{2} + 15.3 g_{n}^{2})
    |0.05925g_{p}^{2} - 0.05830g_{n}^{2}|
   \lesssim 6.4\times 10^{-13}.
\label{final limit}
\end{equation}
A plot of the hyperbolic constraint in Eq.~(\ref{final limit}) is
shown in Fig.~2 along with an illustrative elliptical constraint curve
obtained from Eq.~(\ref{Gundlach's limit}) by substituting 2 = Li and
$2'$ = Ru.  As can be seen from this figure and Eqs.~(\ref{Gundlach's
limit}) and (\ref{final limit}), separate bounds on $g_{p}^{2}$ and
$g_{n}^{2}$ can be inferred by repeating the experiment of Gundlach,
{\em et al.}, with various combinations of appropriately chosen test
masses.  Even though the relevant experiments have not yet been
performed, one can nonetheless obtain useful bounds on $g_{p}^{2}$ and
$g_{n}^{2}$ separately by considering special cases of Eq.~(\ref{final
limit}). For example, for a light pseudoscalar which couples
universally to baryon number we have $g_{p}^{2} = g_{n}^{2}$, and hence
from Eq.~(\ref{final limit})
\begin{equation}
g_{p}^{2}/4\pi \lesssim 4 \times 10^{-7}.
\label{g_p limit}
\end{equation}
The result in Eq.~(\ref{g_p limit}) represents an improvement by more
than two orders of magnitude on the bound inferred by Ramsey
\cite{Ramsey,Adelberger Review}, $g_{p}^{2}/4\pi \lesssim 5 \times
10^{-5}$.  This is the only other direct laboratory limit on
$g_{p}^{2}$, which was obtained by comparing theory and experiment for
the energies of low-lying vibrational and rotational states in
molecular H$_{2}$.  Two other interesting bounds can be inferred from
Eq.~(\ref{final limit}) in the limiting cases $g_{p}^{2}\gg g_{n}^{2}$
and $g_{n}^{2}
\gg g_{p}^{2}$.  These are
\begin{equation}
g_{p}^{2}/4\pi \lesssim 9 \times 10^{-8}, \,\,\,(g_{p}^{2}\gg
g_{n}^{2});\phantom{somespa}
g_{n}^{2}/4\pi \lesssim 7 \times 10^{-8}, \,\,\,(g_{n}^{2}\gg
g_{p}^{2}).
\end{equation}

In contrast to the case for $g_{p}^{2}$, there are no direct laboratory
limits on $g_{n}^{2}$, apart from those arising from Eq.~(\ref{final
limit}).  However, one can attempt to infer a crude indirect bound on
$g_{n}^{2}$ by following an argument due to Daniels and Ni (DN)
\cite{Daniels}.  
Consider, for example, the experiment of Ritter, {\em et al.}
\cite{Ritter}, which uses test samples of Dy$_{6}$Fe$_{23}$ containing
polarized electrons to measure $g_{e}^{2}$.  As noted by DN, the
hyperfine interaction of the electrons in Dy aligns the Dy nuclei and
similarly, but less significantly, for Fe.  DN estimate this
polarization (at room temperature) to be $P_{X} \simeq 3.4
\times 10^{-5}$ ($X = $ Dy), which compares to $P_{e} \simeq 0.4$ for
the electrons themselves.  Hence, although the Dy nuclei have a non-zero
induced polarization, this polarization is quite small.  It follows that
the sensitivity of the experiment of Ritter, {\em et al.} \cite{Ritter}
to
$g_{X}^{2}$ is smaller than its sensitivity to $g_{e}^{2}$ by a factor
$P_{X}^{2}/P_{e}^{2} = 7 \times 10^{-9}$, due to the differences in
$\langle S_{12}\rangle$ for electrons and nuclei.  To infer a bound on
$g_{n}^{2}$ the Dy polarization must be related to that of the
neutron.  If we assume, for example, that the polarization of the Dy
nucleus is carried by a single odd neutron outside a symmetric core,
then we can identify the neutron polarization with that of the Dy
nucleus.  Combining the preceding arguments we are led to the crude
estimate,
\begin{equation}
g_{n}^{2}/4\pi \lesssim (P_{X}^{2}/P_{e}^{2})^{-1}(g_{e}^{2}/4\pi)
 \simeq 8 \times 10^{-6},
\label{Ni's limit}
\end{equation}
where we have used $g_{e}^{2}/4\pi \lesssim 6 \times 10^{-14}$ from
Ritter, {\em et al.} \cite{Ritter}.  Note that although the limits on
$g_{e}^{2}$ from other experiments such as Ref.~\cite{Chiu and
Ni,Wineland} are more restrictive, the configuration  of these
experiments renders the preceding arguments inapplicable
\cite{Long paper}.   In the experiment of Chiu and Ni \cite{Chiu and
Ni}, for example, the polarization of an initially unpolarized
TbF$_{3}$ sample was measured in the presence of a rotating polarized
Dy$_{6}$Fe$_{23}$ source.  Since the TbF$_{3}$ sample was shielded
against conventional magnetic fields by superconducting Nb, any
polarization of the electrons would arise solely from the putative
long-range spin-spin interaction, which is presumably a small effect. 
The alignment of the nuclear spins via the hyperfine interaction would
be smaller still, and hence no useful limit on couplings to nucleons
emerges from such an experiment.

	The laboratory constraints on pseudoscalar couplings derived in
this paper are model independent, but do not apply to axions which
are derivative-coupled \cite{Ferrer}.  Although the present
formalism can be adapted to infer limits on axion couplings using
the $1/r^{5}$ potential arising from 2-axion exchange, the best
existing limits on light axions still come from stellar cooling
\cite{Raffelt,Cheng}.  In addition, astrophysical arguments also yield 
tighter bounds on Yukawa (i.e., non-derivative)
couplings of pseudoscalars to nucleons.  For example, energy loss arguments
from the SN 1987A supernova typically give
$g^{2}/4\pi \lesssim 10^{-21}$ \cite{Raffelt,Grifols 1989}.

In summary, we have shown that the most stringent laboratory limits on
the Yukawa couplings of light pseudoscalars to nucleons (and
ultimately to quarks) derive from the $O(g^{4})$ contributions in
Fig.~1 to equivalence principle experiments.  
These limits  can be further improved by
reconfiguring existing experiments to make them more sensitive to a
short-range
$1/r^{4}$ force, and by using appropriate materials such as Li and Ru. 
Furthermore, by suitably adapting space-based experiments such as STEP
\cite{Reinhard} even more significant improvements in sensitivity could
be realized in the foreseeable future.

The authors wish to thank Jens Gundlach, Wei-Tou Ni, Frank Rickey, and
David Wineland for helpful communications.  This work was supported in
part by the U.S. Department of Energy under Contract No. DE-AC
02-76ER01428.

\begin{figure}
\caption{Contributions to the spin-independent long-range interaction
of fermions $a$ and $b$ arising from two-pseudoscalar-exchange.  The
solid lines are fermions and the dashed lines denote the pseudoscalars.}
\end{figure}

\begin{figure}
\caption{Constraints on $g_{p}^{2}$ and $g_{n}^{2}$ arising from
two-pseudoscalar-exchange.  The region shaded in dark gray exhibits the
hyperbolic constraint implied by the experiment of Gundlach, {\em et
al.}, Ref.~[4].  The light gray region illustrates the
hypothetical elliptical constraint that would emerge from Gundlach, {\em
et al.}, had they used Li and Ru as the test masses.  The overlap
region is shown in black.}
\end{figure}

%\end{multicols}


\begin{references}

\bibitem{Bohr and Mottelson} A. Bohr and B. R. Mottelson, {\em Nuclear
Structure}, Vol. 1 (Benjamin, New York, 1969), p. 249.

\bibitem{Anselm}  A. A. Anselm and N. G. Uralstev, Phys. Lett. {\bf 114B},
39 (1982); {\bf 116B}, 161 (1982).

\bibitem{Gillies} R. C. Ritter, G. T. Gillies, L. I. Winkler, in
{\em Spin in Gravity} ed. by P. G. Bergmann, V. de Sabbata, G. T.
Gillies, and P. I. Pronin (World Scientific, Singapore, 1998),
pp.~199--212.

\bibitem{Gundlach}  J. H. Gundlach, G. L. Smith, E. G. Adelberger, B. R.
Heckel, and H. E. Swanson, Phys. Rev. Lett. {\bf 78}, 2523 (1997).

\bibitem{Fischbach book}  E. Fischbach and C. Talmadge, {\em The Search for
Non-Newtonian Gravity} (AIP Press/Springer-Verlag, New York, 1998).


\bibitem{Ritter}  R. Ritter, C. E. Goldblum, W.-T. Ni, G. T. Gillies, and
C. C. Speake, Phys. Rev. D {\bf 42}, 977 (1990).


\bibitem{Drell}  S. D. Drell and K. Huang, Phys. Rev. {\bf 91}, 1527 (1953);
V. M. Mostepanenko and I. Yu. Sokolov, Sov. J. Nucl. Phys. {\bf 46}, 686
(1987) [Yad. Fiz. {\bf 46}, 1174 (1987)]; 
D. Sudarsky, C. Talmadge, and E.
Fischbach (unpublished).

\bibitem{Ferrer} F. Ferrer and J. A. Grifols,
Phys. Rev. D {\bf 58}, 096006 (1998). 

\bibitem{Long paper} E. Fischbach and D. E. Krause, in preparation.

\bibitem{Sensitivity} The coefficient of $\vec{\cal F}(\vec{r}\,)$ is
proportional to the ``Adelberger sensitivity function''.  See
Ref.\cite{Fischbach book}, and E. G. Adelberger, {\em et al.}, Phys.
Rev. Lett. {\bf 59}, 849 (1987).


\bibitem{Ramsey} N. F. Ramsey, Physica {\bf 96A}, 185 (1979).

\bibitem{Adelberger Review} E. G. Adelberger, B. R. Heckel, C. W.
Stubbs, and W. F. Rogers, Annu. Rev. Nucl. Part. Sci. {\bf 41}, 269
(1991).


\bibitem{Daniels}  J. M. Daniels and W.-T. Ni, Mod. Phys. Lett. A {\bf 6},
659 (1991).




\bibitem{Chiu and Ni} T. C. P. Chiu and W. -T. Ni, Phys. Rev. Lett.
{\bf 71}, 3247 (1993).

\bibitem{Wineland}  D. J. Wineland, J. J. Bollinger, D. J. Heinzen, W.
M. Itano, and M. G. Raizen, Phys. Rev. Lett. {\bf 67}, 1735 (1991).

\bibitem{Raffelt}  G.~G.~Raffelt, Phys. Rep. {\bf 198}, 1 (1990), G. G.
Raffelt, Ann. Rev. of Nucl. Part. Sci. (to be published), hep-ph/9903472.

\bibitem{Cheng} H. -Y. Cheng, Phys. Rep. {\bf 158}, 1 (1988); M. S.
Turner, Phys. Rep. {\bf 197}, 67 (1990).

\bibitem{Grifols 1989}  J. A. Grifols, E. Mass\'{o}, and S. Peris, Mod.
Phys. Lett. A {\bf 4}, 311 (1989).

\bibitem{Reinhard}  R. Reinhard, ed., {\em Proceedings of the STEP
Symposium}, European Space Agency publication ESA WPP-115 (1996).

\end{references}
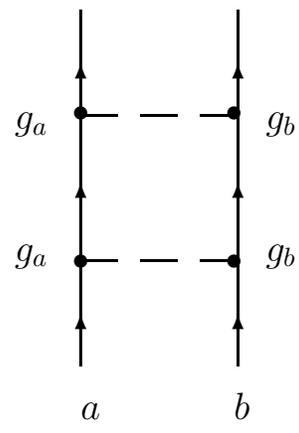
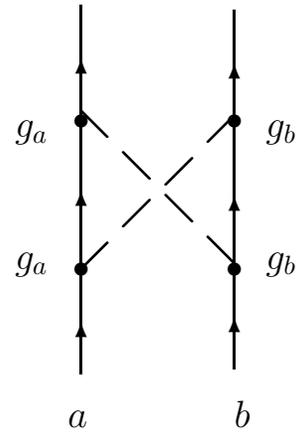
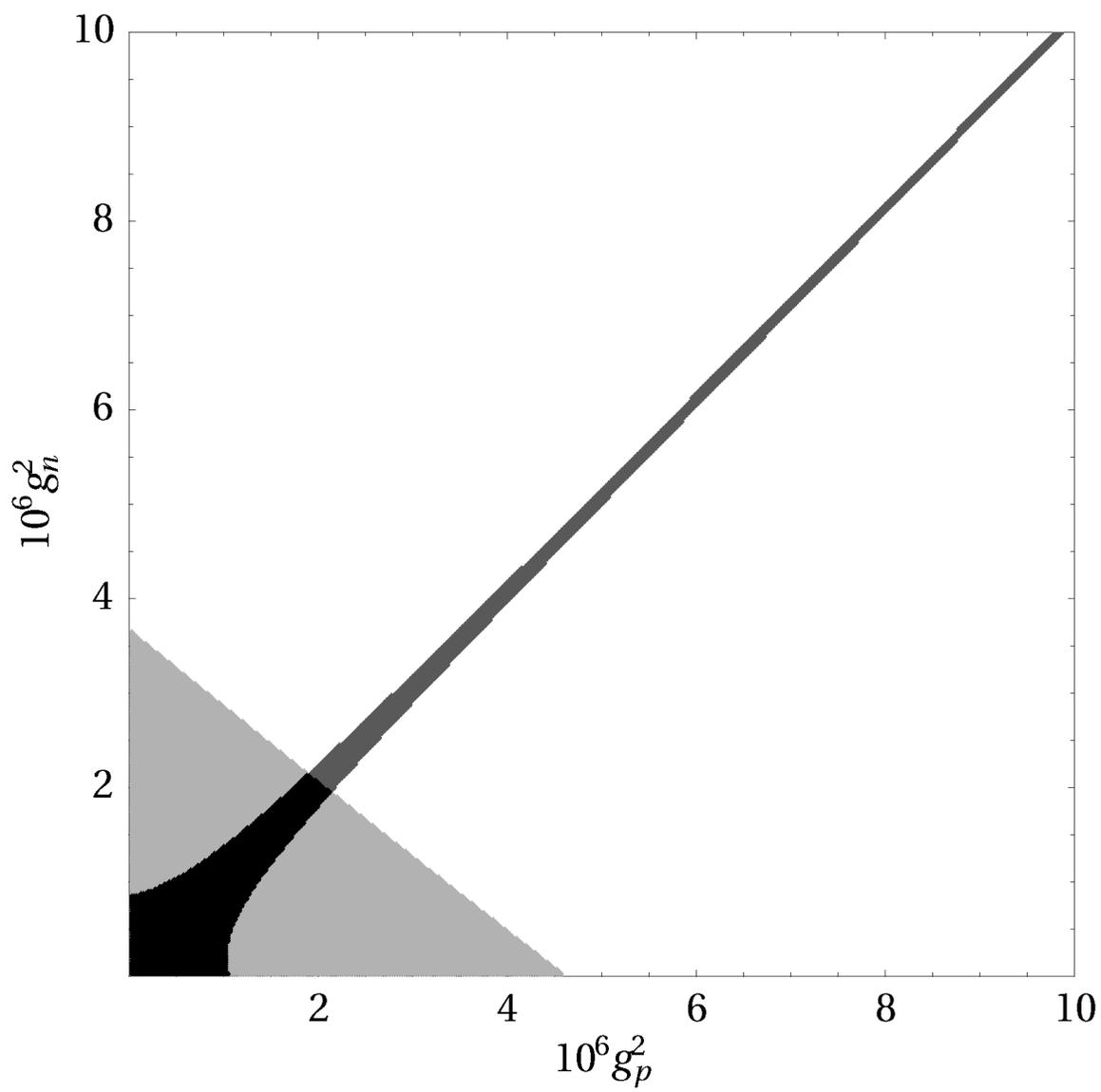
\end{document}